\documentclass[compsoc,conference,a4paper,10pt,times]{IEEEtran}
\IEEEoverridecommandlockouts
\usepackage{cite}
\usepackage{amsmath,amssymb,amsfonts}
\usepackage{algorithmic}
\usepackage{graphicx}
\usepackage{textcomp}
\usepackage{bmpsize}
\usepackage{xcolor}
\usepackage{lipsum}
\usepackage[colorlinks=true,urlcolor=black]{hyperref}
\def\BibTeX{{\rm B\kern-.05em{\sc i\kern-.025em b}\kern-.08em
    T\kern-.1667em\lower.7ex\hbox{E}\kern-.125emX}}

\usepackage{amsthm}
\usepackage{caption}
\usepackage{moeptikz}
\usepackage{url,breakurl} 
\usepackage{longtable}
\usepackage{enumitem}
\usepackage{array}
\usepackage{subfigure, wrapfig}
\usepackage{tikz}
\usepackage{epsdice}
\usepackage{listings}
\usetikzlibrary{shapes,arrows,positioning,calc}
\usetikzlibrary{arrows.meta} 

\definecolor{blue}{HTML}{008ED7}
\definecolor{lightBlue}{HTML}{e5f7ff}

\title{Echomix: a Strong Anonymity System with Messaging}

\author{
\IEEEauthorblockN{Ewa J. Infeld}
\and
\IEEEauthorblockN{David Stainton}
\and
\IEEEauthorblockN{Leif Ryge}
\and
\IEEEauthorblockN{Threebit Hacker}
}

\date{October 2024}

\begin{document}

\maketitle

\begin{abstract}
Echomix is a practical mix network framework and a suite of associated protocols providing strong metadata privacy against realistic modern adversaries. It is distinguished from other anonymity systems by a resistance to traffic analysis by global adversaries, compromised contacts and network infrastructure, quantum decryption, and statistical and confirmation attacks typical for multi-client messaging setting. It is implemented as Katzenpost, a robust software project, and used in multiple deployed systems, and features relatively low latency and bandwidth overhead.

The contributions of this paper are: (1) Improvements on leading mix network designs, supported by rigorous analysis. These include solutions to crucial vulnerabilities to traffic analysis, malicious servers and active attacks. (2) A cryptographic group messaging protocol with strong metadata protection guarantees and reliability. 
(3) Hybrid post-quantum nested packet encryption. 

\end{abstract}


%

\section{Introduction}



Protecting metadata is as crucial a concern for privacy as protecting the content of communications. They are a primary stock of data brokers \cite{broker} and qualifier in large dataset analysis. Surveillance actors take advantage of metadata \cite{metadata, belarus, fellowtraveller} even to exert lethal force. \cite{killmetadata, whatsapp} 

Anonymity systems in use today offer incomplete protections against the most powerful class of surveillance adversaries. The research field of anonymity is robust, yet many academic designs disregard real-world Internet conditions, or explicitly declare as non-goals the resistance to a wide variety of practical attacks on user data and metadata. Second-party anonymity is seldom considered. 

We describe a novel, implemented, practical and reliable mix network design, with a threat model that allows for sophisticated, global, active adversaries who may compromise network elements and a user's contacts, have access to a quantum computer, and do powerful cryptanalysis. 
We overcome weaknesses of leading anonymity systems, and introduce protocols which provide strong security guarantees even in the case of persistent messaging between users. The design and software is used in a growing number of deployed systems, \cite{0kn, cloaked}
as well as our own chat client. \cite{katzen}

This is in contrast to systems such as Tor \cite{tor}, 
which does not protect against a global adversary \cite{torgpa1, torgpa2, torgpa3} and is vulnerable to an array of confirmation and traffic attacks \cite{torsurvey, torconfirmation1}, some of which have been exploited by surveillance actors \cite{torrelatorNDR}. The leading mix network model today is Loopix \cite{loopix}, which is a basis of Nym's system \cite{nym}. Our design eliminates its many shortcomings, including vulnerability to traffic analysis, receiver observability, and vulnerability to malicious service providers. We support these claims with rigorous analysis. 

We additionally introduce a messaging protocol which is suitable for anonymous group messaging with a realistic threat model, and provides reliability without forcing interactivity. We then present a quantum resistant packet format appropriate for mix networks. Finally, we provide latency and bandwidth overhead evaluation, to demonstrate that this system is practical. 




\section{Threat model}

We consider a realistic modern adversary, such as a government surveillance agency, a large technology corporation, or a criminal organization. The adversary is:

\newtheorem*{Global}{Global}
\begin{Global} \normalfont The adversary can see all or a significant portion of connections of the entire global internet and is capable of statistical analysis of gathered data. 
For many attacks that are typically attributed to a global adversary, it is enough if the adversary has a view of a target population of users of the network. 

\end{Global}

\newtheorem*{Active}{Active}
\begin{Active}\normalfont
The adversary can disable parts of the network, and plant or take over some devices in the network to inject malicious code and gain access to the information available to them. This can happen by technical means, exercising legal or extralegal forms of coercion, or subterfuge. 
The adversary can compromise a client's contacts' devices, resulting in a need for \emph{second-party anonymity} in the system. We refer to a compromised contact as a second-party adversary, or 2PA. We minimize metadata shared with contacts and avoid various forms of forced interactivity such as automatic delivery and read receipts. 

Many attacks typically attributed to an active adversary are also a concern with a passive adversary able to observe network disruption events. If a user's microwave oven turns on and causes a brief connection disruption for their WiFi, that event can be visible to various unrelated internet services which know the user's identity. In particular, connection disruptions must not be revealed to contacts.\footnote{These network disruption confirmation attacks have been used by surveillance actors. \cite{hammond}, \cite{torrelatorNDR} In particular, \cite{hammond} is an example of a 2PA.}

Echomix is not secure against an active adversary who compromises the entire system, or close to the entire system, such as a majority of directory authorities or a critical combination of node types on the client's path, as described in \autoref{pigeonhole-security}.
\end{Active}

\newtheorem*{Sophisticated}{Sophisticated}
\begin{Sophisticated}\normalfont
The adversary has large computational resources, and is capable of cryptanalysis on par with frontier research. The adversary has access to a quantum computer, or will have access to one in the near future.
\end{Sophisticated}

\newtheorem*{Context}{Has context}
\begin{Context}\normalfont
The adversary can supplement collected data with rich context of already gathered data on all users from other sources.
\end{Context}



\noindent We will define \textit{strong anonymity} for classification purposes as an ability to withstand a sophisticated, \textbf{global passive adversary (GPA)}. As demonstrated, the Echomix threat model goes further to include active adversaries of significant, but not absolute, power.

We define metadata broadly, to include all observable distinguishing characteristics of a user's activity. This can be sender and receiver identity, timing information, volume and type of communication, social network etc.


\subsection{Anonymity systems in light of the threat}

Several existing types of communication systems offer anonymity protections. We survey these briefly and note weaknesses with respect to our stated threat model. 

\subsubsection{VPNs} VPNs relay a user's connection while minimizing latency and maximizing bandwidth. Traffic can be correlated by a passive adversary observing network traffic at the the VPN, or the source and destination traffic.

Some VPNs are overlayed on mix networks and/or employ decoy traffic \cite{nymvpn
, cloakedvpn
}. The strongest of these designs attempt to transport VPN traffic with uniform bandwidth by clients, necessitating both a large bandwidth overhead and an upper bandwidth limit. However, Internet protocols facilitated by VPNs allow for confirmation attacks through forced interactivity, and so interruptions to client traffic can be correlated, either by passive observation 
or active interruptions of clients.

\subsubsection{Tor}
Tor is much more sophisticated than a VPN, and does not rely on a single or few points of failure. In the absence of a GPA, we consider Tor to be state-of-the-art for practical anonymous communication and it is able to provide users with an experience of browsing the Internet comfortably. However, a passive observer able to see two endpoints can correlate connections \cite{torrelatorNDR, torgpa1, torgpa2, torgpa3}. As with VPNs, because Tor is a general purpose tool used to transport protocols which force interactivity, we consider it impossible to extend Tor's threat model to include protection against a GPA. \cite{torsurvey, torconfirmation1}

\subsubsection{Mix networks}

Mix networks, or \emph{mixnets}, are an anonymous communications network paradigm distinguished by striking a balance between practicality and protection from strong adversaries. 
A mix network consists of network devices, typically referred to as mix nodes, that relay messages between clients in such a way that an adversary is unable to determine which clients are communicating with each other. This is done by reordering or mixing multiple indistinguishable messages. 

There must be some trade-off in latency - packets must not be forwarded from a node immediately, in order to be \textit{mixed} with other packets, and some trade-off in bandwidth - packets must be padded to a common size. They also have to be re-encrypted at each hop. \cite{chaum} The notion that both latency and bandwidth overhead must be non-zero was formalized in \cite{trilemma}. Additional trade-offs occur with adding common anonymity strategies, such as decoy traffic. It is thanks to these combined strategies and trade-offs that carefully constructed mix networks can offer protection against global adversaries. Because of these trade-offs mix networks are not able to simulate browsing the Internet comfortably as a VPN might, and are typically considered for use with some services, such as messaging.

Many mix network designs and protocols have been published, starting in 1979 \cite{chaum}, and we observe a rise in commercial efforts to build mixnets \cite{nym, 0kn,hopr}. Much of the focus of these efforts is in incentivisation mechanisms for mix node operators, and financial privacy. The largest of these, Nym \cite{nym}, is based on the Loopix model. \cite{loopix} Due to a subtle oversight in design, Loopix is vulnerable to a GPA as described in \autoref{section-trafficanalysis}. It also has a single point of failure in a user's Service Provider, making it easy for an active adversary to compromise a target user. 



\subsubsection{Non-mixnet theoretical systems with strong anonymity}
There exist theoretical designs providing strong anonymity, some of which have information-theoretic guarantees, which Echomix does not. They all carry significant overheads which have so far made them impractical for real world deployment at a scale. These include DC-nets and other $k$-anonymity-based systems \cite{kanonymity, Bleumer2011, buddies, carnivores}, as well as PIR \cite{talek} designs. 
 



\section{The Echomix design}

The Echomix system contains three server node types: gateways, mix nodes and services. None of these has a persistent relationship with clients. Service nodes are positioned on the far side of the network, and, in contrast to leading mix network designs, every interaction with them is a round trip - an \emph{echo}.  When a client connects to a random gateway, the Sphinx \cite{sphinx} packets sent by the client are relayed through the three layers of mix nodes, and then to a service. A reply is sent, which can confirm delivery or contain send query results. A service can send this reply without knowing the location of the client thanks to a Sphinx Single-Use Reply Block (SURB). 

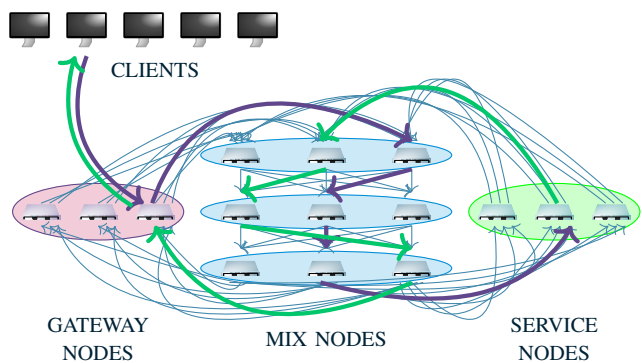
\begin{figure}[!ht]
\centering
\begin{tikzpicture}[scale=0.75]

\filldraw[opacity=0.1,purple!20!white] (-2,1.25) ellipse (1.5cm and 0.5cm);
\filldraw[opacity=0.1,green!20!white] (6,1.25) ellipse (1.5cm and 0.5cm);
\filldraw[opacity=0.1,blue!20!white] (2,0.25) ellipse (2.2cm and 0.3cm);
\filldraw[opacity=0.1,blue!20!white] (2,1.25) ellipse (2.2cm and 0.3cm);
\filldraw[opacity=0.1,blue!20!white] (2,2.25) ellipse (2.2cm and 0.3cm);
\draw[opacity=0.6,purple!50!blue] (-2,1.25) ellipse (1.5cm and 0.5cm);
\draw[opacity=0.6,green] (6,1.25) ellipse (1.5cm and 0.5cm);
\draw[opacity=0.6,blue] (2,0.25) ellipse (2.2cm and 0.3cm);
\draw[opacity=0.6,blue] (2,1.25) ellipse (2.2cm and 0.3cm);
\draw[opacity=0.6,blue] (2,2.25) ellipse (2.2cm and 0.3cm);

\foreach \N in {0,1.5,3}{
\foreach \L in {0,1,2}{
\draw [->,blue!50!gray,opacity=0.35] plot [smooth, tension=1] coordinates { (-1-\L+\N*0.1,1.4) (0+\N*0.2,2.5+\N*0.2) (\N+0.4+0.1*\L,2.5)};
\draw [->,blue!50!gray,opacity=0.35] plot [smooth, tension=1] coordinates { (4.9+\L+\N*0.1,1.4) (3.5+\N*0.3-0.1*\L,3.25+\N*0.1) (\N+0.3+0.1*\L,2.5)};
\draw [->,blue!50!gray,opacity=0.35] plot [smooth, tension=1] coordinates { (\N+0.5+\L*0.1,0) (-\N*0.1-0.2*\L,-\N*0.2+\L*0.1) (-1-\L+\N*0.1,1)};
\draw [->,blue!50!gray,opacity=0.35] plot [smooth, tension=1] coordinates { (\N+0.5-\L*0.1,0) (5-\N*0.1-0.2*\L,-\N*0.2+\L*0.1) (5.25+\L-\N*0.1,1)};}
}
\foreach \x in {0.8,-0.2,-1.2,-2.2,-3.2}{
\node[client, minimum size=0.5cm] at (\x,4.5) {};}
\foreach \L in {-0,1,2}{
\node[server, fill=purple!50!blue, minimum size=0.5cm] at (-\L-1,1.25) {};
\node[server, fill=cyan!50!black, minimum size=0.5cm] at (\L+5,1.25) {};
\foreach \N in {0,1.5,3}{
\node[server, fill=blue!50!black, minimum size=0.5cm] at (\N+0.5,\L+0.25) {};
}}
\foreach \L in {0,1}{
\foreach \N in {0,1.5,3}{
\foreach \k in {0,1.5,3}{
\draw[->, blue!50!gray, opacity=0.35] (\N+0.5,\L+1) -- (\k+0.5,\L+0.6);
}}}
\draw[blue!25!black] (-1,3.75) node {\footnotesize  CLIENTS};

\draw[blue!25!black] (-2,-0.75) node {\footnotesize GATEWAY};
\draw[blue!25!black] (-2,-1.25) node {\footnotesize NODES};
\draw[blue!25!black] (2,-1) node {\footnotesize  MIX NODES};
\draw[blue!25!black] (6,-0.75) node {\footnotesize SERVICE};
\draw[blue!25!black] (6,-1.25) node {\footnotesize NODES};
\draw [->,blue!50!purple,ultra thick,opacity=1] plot [smooth, tension=1] coordinates { (-0.75-3*0.1,1.4) (0+3*0.2,2.5+3*0.2) (3+0.4+0.1,2.5)};
\draw[->, purple!50!blue, ultra thick, opacity=1] (3.5,2) -- (2,1.6);
\draw[->, blue!50!purple, ultra thick, opacity=1] (2,1) -- (2,0.6);
\draw [->,blue!50!purple, ultra thick, opacity=1] plot [smooth, tension=1] coordinates { 
(2-0.1,0) (5-0.15-0.2,-1.5*0.2+0.1) (4.9+1.5-0.1,1)};

\draw [->,green!50!blue,ultra thick,opacity=1] plot [smooth, tension=1] coordinates { 
(6.05,1.4) (3.6+0.45,3.25+0.15) (1.6+0.3,2.5)};
\draw[->, green!50!blue, ultra thick, opacity=1] (2,2) -- (0.5,1.6);
\draw[->, blue!50!green, ultra thick, opacity=1] (0.5,1) -- (3.5,0.6);
\draw [->,blue!50!green,ultra thick,opacity=1] plot [smooth, tension=1] coordinates { (3.5,0) (0.75,-0.5) (-0.8-2*0.1,1)};

\draw [->,blue!50!purple,ultra thick,opacity=1] plot [smooth, tension=1] coordinates { (-2.2,4) (-2.25,2.5) (-1.2,1.4)};
\draw [->,green!50!blue,ultra thick,opacity=1] plot [smooth, tension=1] coordinates { (-1.4,1.35) (-2.4,2.5) (-2.4,4)};

\end{tikzpicture}
\caption{A client's interaction with the service node is a round-trip, with a packet's forward route marked in purple, and service's confirmation in green. The intermediate node layers are pictured from top to bottom.}
\end{figure}
\vskip -0.1 in 

In the context of a messaging application, a message travelling from Alice to Bob requires two \emph{echos} to the far side of the network, one from Alice to write a message to a service, and a second initiated by Bob to retrieve it. This symmetry and a careful construction of protocols, allows us to out-perform other anonymous communication systems in resisting traffic analysis, malicious providers, and confirmation attacks.

The clients generate a stream of \emph{echos} to uniformly random, or pseudorandom - in the case of some applications, including message streams - services, independently of whether a client requests a service or not. Most of these packets will be decoy traffic. The stream is a single Poisson process, with delays at each hop sampled from an exponential distribution, as in \cite{stopandgo} and \cite{loopix}. The advantages of this strategy are described in section \ref{section-memoryless}. Unlike in Loopix \cite{loopix}, the amount of sent and received traffic is fully independent of whether the packets are application traffic or decoys.


 From the point of view of network architecture, Echomix, and its implementation - Katzenpost \cite{kp}, is an Internet overlay mix network atop TCP/IP or QUIC/IP. The subsequent layers are governed by the following protocols.

\begin{enumerate}[itemsep=0ex, parsep=0ex, parsep=0ex]
    \item PQ Noise \cite{pqnoise} transport protocol.
    \item Sphinx \cite{sphinx} or PQ-Sphinx routing protocol.
    \item Application layer, such as Pigeonhole messaging.
\end{enumerate}

Katzenpost is the first software implementation of the PQ Noise. As a transport protocol it enforces the network topology, e.g. mix nodes in layer 1 are only allowed to downstream-connect to the mix nodes in layer 2. 
All PQ Noise messages are padded to a uniform size.

Since all application protocols are built based on packet round-trips to services positioned on the far side of the network, the Sphinx ability for the service to send a reply without knowing the location of the client is used throughout the design. This is done with Single Use Reply Blocks (SURBs). We describe our post quantum updates to the Sphinx format in \autoref{section-pq}.

The Katzenpost PKI is an adjacent protocol at the root of authority within the mix network. Directory authorities publish PKI documents every epoch, distributing network connection information and public key materials to the nodes. This system is similar to Tor's. \cite{tordirspec} Providing all nodes and clients with a uniform view of the network allows us to resist epistemic attacks. \cite{ep1} An adversary who compromises a majority of directory authorities compromises the entire mix network. We therefore employ a decentralised PKI design, relying on multiple independent directory authority operators. The consensus producing the PKI document uses the post quantum hybrid signature scheme of Ed25519 \cite{ed25519} combined with Sphincs+ \cite{sphincs}.

\subsection{Gateway nodes}

In contrast to \cite{loopix}, we believe that the nodes on the edge of the network should have as little information on the client behavior as possible, as they are the ones that can identify the client. Gateways have no persistent relationship with the client, and no knowledge of whether any of the client's packets are decoys or not.

If the client traffic crossing the node is low, the gateways additionally generate decoy traffic. We propose a new heuristic, the \textit{Coupon Collector's bound} in order to ensure that all links of the network are active, resulting in a reliable mixing of packets.

\newtheorem*{CCB}{Coupon Collector's Bound}
\begin{CCB}
If packets released from a node are directed to one of the nodes in the next layer uniformly at random, the expected number of messages that node has to release before at least one message has been directed to each node in the subsequent layer behaves like  $\Theta (n \log (n) )$. \medskip

\noindent\textit{Proof:} This is a direct consequence of the \textit{Coupon Collector's Problem}.\hfill $\Box$
\end{CCB}
\noindent Let:
\begin{itemize}[itemsep=0ex, parsep=0ex, parsep=0ex]
    \item $\mu=1/\lambda$ be the mean time that a packet lingers in a node due to memoryless mixing,
    \item $n$ be the maximum number of nodes in a layer,
    \item $g$ be the number of gateways.
\end{itemize}
We aim to achieve a high probability that during each period $\mu$, there is at least one packet crossing a given link in the network. The Coupon Collector's Problem tells us how many packets we need to cover links from the gateways to the first layer of mixes. In order for all links between two layers of size $n$ to be reliably active, we should additionally multiply the desired output of a gateway by $n/g$. Therefore in order for all links to be active with high probability, and therefore for best mixing of the packets in the network, each gateway should be outputting at least an average of $\Theta (n^2 \log (n)/g )$ packets in the time $\mu$. In practice, this bound is significantly lower than the number of packets travelling through the nodes.

\subsection{Decoy traffic and application traffic}\label{section-coupling}

An \emph{echo} decoy packet is a Sphinx round-trip through the mix network to a service node sampled uniformly at random. A service request to a server sampled in a way indistinguishable from uniformly random, followed by a service response sent back with the SURB, is unobservably coupled to echo decoy traffic.

\newtheorem*{COP}{Traffic coupling}\begin{COP}Let sequences $A$ and $B$ be two probabilistic processes on states $S=\{s_1,s_2,\dots,s_n\}$. If both $A$ and $B$ are indistinguishable from uniformly random, then any process $C$ which selects either $A$ or $B$ to sample the next state according to any algorithm $\mathcal{A}$ is also indistinguishable from uniformly random. \medskip

\noindent Proof: Let $\mathcal{H}$ be any state history in $C$, and $s_j\in S$ be any state. Then for any values of $p=\mathbb{P_\mathcal{A}]}(A|\mathcal{H})$ and $p'=\mathbb{P_\mathcal{A}}(B|\mathcal{H})$,the probability of the next element being from sequence $A$ and sequence $B$ respectively, the probability of the next sampled state being $s_j$ is $$p\times \mathbb{P}_A(s_j|\mathcal{H})+p'\times \mathbb{P}_B(s_j|\mathcal{H})=(p+p')\times \frac{1}{n}=\frac{1}{n}. $$$ \mathcal{A}$ does not have to be independent of history $\mathcal{H}$. \hfill $\Box$.
\end{COP}

\noindent If and only if we maintain the correct coupling, the user's application traffic is fully unobservable within decoy traffic, and causes no anomalies, avoiding the shortcomings of leading systems described in \autoref{section-trafficanalysis}. 
For receiving messages, maintaining unobservability requires care, since not only are the queries correlated with the related send requests, but we may have to contend with the dangers of multiple queries to the same server, which may be statistically significant and therefore visible to a passive observer. The introduction of \emph{couriers} (\autoref{pigeonhole-section}) allows us to efficiently maintain unobservably coupled traffic. It also prevents \emph{SURB floods} - an attack in which a malicious service 
might collect multiple linkable SURBs and send them in a burst in order to identify a client. 

\subsection{Service nodes}\label{service-nodes}

Service nodes are positioned behind the mix network and handle functionality requested by the client. This could be storing messages, publishing information outside of the mixnet, interfacing with a blockchain node etc. They also process and execute the SURBs of decoy packets. Any application should be constructed so that the following conditions are met:
\begin{enumerate}[itemsep=0ex, parsep=0ex, parsep=0ex]
\item Separate service requests of a client are unlinkable. Repeating the same request may be linkable.
\item Services are treated uniformly by the client, with no persistent relationship.
\item The traffic from a client to the service is correctly coupled with the decoy traffic. 
\end{enumerate}

The last condition means that either the service is chosen in a way that is independent from traffic history and indistinguishable from uniformly random for each query to a service, or the packet will replace a decoy packet that was meant to go to the specific service. Since the latter introduces additional latency, we achieve the former in several ways in sections \ref{section-bacap} and \ref{pigeonhole-section}.


\subsection{Iterating on prior research}

So far, we described original additions to the field of mix network design. For the remainder of this section, we will discuss how we implement and improve upon established concepts, and motivate the new design by pointing out vulnerabilities of preceding systems. Beginning with \autoref{section-bacap}, we list remaining original contributions which allow us to extend our security guarantees to persistent messaging, and resisting quantum adversaries.

Katzenpost grew out of the Panoramix project \cite{panoramix}, and was previously an evolution of Loopix \cite{loopix}, which combined ideas of sampling an exponential distribution (memoryless mixing) to delay at each hop \cite{stopandgo}, monitoring system health with heartbeat traffic \cite{heartbeat}, organizing nodes in a layered topology \cite{topology}, assigning persistent Service Providers for each user at the edge of the network, and wrapping data in Sphinx packets \cite{sphinx}, a packet format designed specifically for mix networks.






 
\subsubsection{Memoryless mixing}\label{section-memoryless} Echomix adopts node delays sampled from an exponential distribution, as introduced in \cite{stopandgo} and used in Loopix \cite{loopix}. This distribution has the advantage of being \emph{memoryless} - at each point in time, each message sitting in a mix will have the same probability distribution of the remaining delay, independently of how long it has already been waiting. This means that for an external observer the probability distribution of which message will be sent next is uniform at all times. For a constant parameter $\lambda >0$ with the mean $1/\lambda$, the delays approximate the probability distribution function: $$f(x_{\geq 0},\lambda)=\lambda e^{-\lambda x} .$$ However, \cite{stopandgo} claims incorrectly that the behavior of a resulting mix node is Poisson distributed. This is then reproduced in \cite{loopix}, culminating in nodes of these type being called \emph{Poisson mix}es. The behavior of a sum of multiple Poisson processes is in fact only Poisson if the set of processes being summed doesn't change over time. We propose to call this type of mixing \emph{memoryless mixing} instead, while noting that the overall behavior of the node itself is not memoryless and a node should not be called a \textit{memoryless mix}.

\subsubsection{Heartbeat packets}

All nodes in the mixnet generate \emph{heartbeat packets} \cite{heartbeat}, which are sent through the system before returning to their origin. This allows each node to take stock of the functioning of the network - if any segment of the route is disrupted, packets crossing through that segment will not come back in the expected time window. This allows the system to detect an $n-1$ attack \cite{n-1}, or any other attack that involves a disruption of the network.
However, heartbeat packets not originating from the clients are only effective in detecting faulty or malicious behavior in some mix nodes. Malicious nodes on the edge of the network can distinguish between mix heartbeats and user requests and decide to forward any heartbeat packet that is scheduled to return to a mix node, while dropping user traffic meant to exit the network. 

Additionally, no node or client in the network should act on these measurements by itself. Such a process could result in different clients having different views of the network and opening the system to \emph{epistemic} \cite{ep1} and \textit{compulsion} \cite{compulsion} attacks. Instead, nodes in Echomix process the collected measurements \cite{katzenpki} 
and rate the links' health, and then upload the ratings to directory authorities. The directory authorities establish a consensus and distribute the updated structure of the network to all parties in the next PKI document, once per consensus epoch.

Finally, both \cite{heartbeat} and \cite{loopix} describe rating nodes in the mixnet. It is more helpful to rate links on the route, rather than nodes. Not only does it provide finer data, and allow for detection of $n-1$ attacks directly, but in practice many networking problems happen specifically between two locations. A separate argument in favor of indexing over links can be found in \cite{miranda} for batch mixes.

\subsubsection{Vulnerability to traffic analysis in Loopix}\label{section-trafficanalysis}


This Loopix design vulnerability has not been addressed elsewhere. A legitimate message traveling from the last layer of mixes is sure to go to the receiver's designated provider, as opposed to decoy traffic from the last layer of mixes, which is uniformly distributed among providers. Application traffic at the last hop is therefore independently overlayed on decoy traffic. 

\begin{figure}[!ht]
\centering
\begin{tikzpicture}[scale=1]

\draw[opacity=0.8,purple!50!blue] (-1,1.25) ellipse (0.6cm and 1.5cm);
\filldraw[opacity=0.1,purple!20!white] (-1,1.25) ellipse (0.6cm and 1.5cm);
\foreach \x in {0,1,2}{
\draw[opacity=0.9,blue] (2,\x+0.25) ellipse (2.2cm and 0.4cm);
\filldraw[opacity=0.1,blue!20!white] (2,\x+0.25) ellipse (2.2cm and 0.4cm);}

\draw[purple!20!black] (-1,3) node {\footnotesize PROVIDERS};
\draw[blue!40!black] (-2.5,1.5) node {\footnotesize ALICE};
\draw[blue!40!black] (-2.5,0.05) node {\footnotesize BOB};

\draw[blue!60!black] (4.75,2.35) node {\footnotesize MIX};
\draw[blue!60!black] (4.75,2) node {\footnotesize LAYER 1};
\draw[blue!60!black] (4.75,1.35) node {\footnotesize MIX};
\draw[blue!60!black] (4.75,1) node {\footnotesize LAYER 2};
\draw[blue!60!black] (4.75,0.35) node {\footnotesize MIX};
\draw[blue!60!black] (4.75,0) node {\footnotesize LAYER 3};

\draw [->,black!50!blue,ultra thick] plot [smooth, tension=1] coordinates { (-2.2,2) (-2,1.5)(-1.5,1.25)};
\draw [->,black!50!blue,thick,opacity=1] plot [smooth, tension=1] coordinates { (-1.4,0.3) (-1.8,0.7) (-2.1,0.45)};

\foreach \N in {0,1.5,3}{
\draw [->,blue!50!gray,opacity=1] plot [smooth, tension=1] coordinates { (-0.75-\N*0.1,1.2) (0+\N*0.2,2.5+\N*0.2) (\N+0.5,2.5)};
\foreach \L in {0,1,2}{
\draw [->,blue!50!gray,opacity=0.5] plot [smooth, tension=1] coordinates { (\N+0.5+\L*0.1,0) (-\N*0.1-0.2*\L,-\N*0.2+\L*0.5) (-0.8-\N*0.1,\L+0.1)};}
}

\node[client, minimum size=0.5cm] at (-2.5,0.5) {};
\node[client, minimum size=0.5cm] at (-2.5,2) {};

\foreach \L in {-0,1,2}{
\node[server, fill=purple!50!blue, minimum size=0.85cm] at (-1,\L+0.25) {};
\foreach \N in {0,1.5,3}{
\node[server, fill=blue!50!black, minimum size=0.85cm] at (\N+0.5,\L+0.25) {};
}}
\foreach \L in {0,1}{
\foreach \N in {0,1.5,3}{
\foreach \k in {0,1.5,3}{
\draw[->, blue!50!gray, opacity=0.5] (\N+0.5,\L+1) -- (\k+0.5,\L+0.6);
}}}

\draw [->,blue!50!black,opacity=1] plot [smooth, tension=1] coordinates { (3+0.5+0.1,0) (-3*0.1,-3*0.2) (-0.8-3*0.1,0.1)};
\end{tikzpicture}
\caption{In Loopix, as Alice communicates with Bob, 
the increase in traffic to Bob's Provider is observable. 
}
\end{figure}
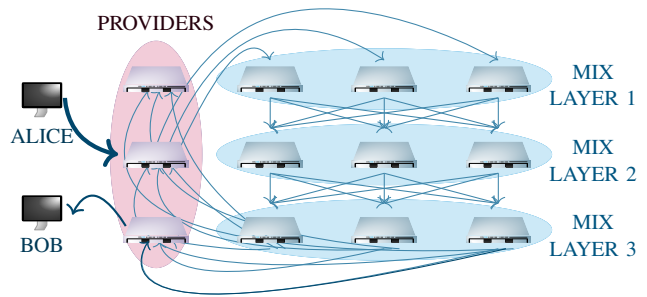

This means that application traffic at that hop is observable to a passive network adversary, as long as it is statistically significant, and especially if there is any regularity to it. In this situation, the other traffic at this hop is not an effective cover, it's noise, and in many real-life situations a signal processing analysis could easily do away with it. This is dangerous even in the short term, since the low latency between the sender and the receiver's Provider opens this system to correlation attacks and SURB floods, and any advantage that comes from asynchronicity is lost. 


 A naive solution if one wanted to retain the rest of the Loopix design might be to wait until a decoy packet is going to the right provider, and send the message instead. This would significantly reduce the available bandwidth by effectively dividing it by the number of Providers. In sections \ref{section-bacap} and \ref{pigeonhole-section}, we carefully designe a correct coupling of decoy and application traffic which does not reduce the bandwidth. 

\subsubsection{Providers}

Loopix features persistent \emph{Service Providers} that the clients connect to directly. The Provider is a significant point of failure, with access to a trove of a user's information, including the message receiving pattern. It was suggested in \cite{loopix} that returning heartbeat traffic may double as decoy traffic for received messages, but this is incorrect. These streams are independently overlayed, and can be similarly decoupled as above when a client is online. When the receiver is not online, loop traffic is not present at all. In Echomix we only include gateways on the perimeter of the network, with service providers positioned behind it, and do not include persistent providers or gateways for a user at all. Interactions of the same user are unlinkable by the service provider.

\subsection{The challenges of persistent multi-client messaging}

Persistent messaging between multiple clients comes with an additional set of statistical and active attacks, and a need for second party anonymity. The design described so far is appropriate for a wide array of applications, including one-to-all anonymous publishing, interfacing with blockchain nodes, and other sender-only use cases. It is a significant improvement on the security properties of previously published mixnet designs.
In order to further extend it to persistent messaging, we introduce the BACAP and Pigeonhole suite of protocols to allow for messaging while maintaining our security guarantees. These are described in sections \ref{section-bacap} and \ref{pigeonhole-section}.

\section{BACAP}\label{section-bacap}

We have established in \autoref{service-nodes} that we require services to be accessed by clients uniformly at random, or in a way indistinguishable from uniformly random. We also require multiple interactions of the same client to be unlinkable by a service. But if users are to be able to exchange messages, they need to know where from and how to retrieve them. Our goal is therefore to implement a form of \emph{private distributed hash table}, wherein users can leave each other messages in pseudorandom locations, with servers able to verify the validity of the message, and read ability, without being able to determine that two messages belong to the same conversation.

BACAP (Blinding-and-Capability scheme) allows us to deterministically derive a sequence of key pairs using blinding, built upon Ed25519 \cite{ed25519}, and suitable for unlinkable messaging. 
It enables participants to 
derive \emph{box ID}s and corresponding encryption keys for independent, single-use \emph{boxes} using shared symmetric keys. 

A box consists of an ID, a message payload, and a signature over the payload. There are two basic \emph{capabilities} - one that lets a party derive the box IDs and decrypt the messages, and one that additionally lets the holder derive private keys to sign the messages. The signatures are universally verifiable, as the box ID for each box doubles as the public key for the signatures. 

In the context of a messaging system, the protocol is used by Alice to send an infinite sequence of messages to Bob, one per \emph{box}, with Bob using a separate, second instance of the protocol to send messages to Alice. Alice will use a root private key to derive a root public key shared between participants. The root key and a CSPRNG instantiated from recursive KDF applications are then used to obtain a sequence of context-specific values for exercising and verifying a capability. A context value $\textrm{ctx}$, which is a hash of a universally public value, will be used as additional input. It can, for simplicity, be a hash of the name of the storage network, or can be bound to a specific period of time, e.g., the \emph{long epoch SRV} published by the Echomix directories at regular intervals, similar to how Tor uses its \emph{SRV} in \cite{tor-v3-onion-services}. The context value makes it safe to \emph{unlinkably} relocate messages to a different network.

\medskip

\noindent All parties (and adversaries) know public constants:

\begin{itemize}[itemsep=0ex, parsep=0ex, parsep=0ex]
    \item $B$ the ed25519 base point
    \item $\ell$ the prime defined in \cite{ed25519}
    \item $\textrm{ctx}$ hashed network context
\end{itemize}

\noindent In the section below we will use the syntax $B \cdot x$ to denote "scalar multiplication" of the point $B$ and scalar $x$, and $a \times b$ to denote natural number multiplication.  The following values are generated by Alice and constitute her \emph{"write capability"} for the sequence:

\begin{itemize}[itemsep=0ex, parsep=0ex, parsep=0ex]
    \item $S_R\in \mathbb{Z}_\ell $: root private key,
    \item 
    $P_R = B \cdot S_R$: root public key.
    \item $i_0 \in \mathbb{Z}_{2^{63}}:$ random initial index counter. 
    \item $H_{i_0}\in \mathbb{Z}_{2^{256}}:$ random initial KDF state.
\end{itemize}

\noindent Alice sends the following \emph{read capability} to Bob out-of-band:
$$P_R,\ H_{i_0},\ i_0. $$
\noindent Our $i$ will be encoded as an unsigned 64-bit integer, and serves to define an ordering for boxes and to enable applications to refer to boxes uniquely. Initializing it with an upper bound of $2^{63}$ ensures that a sequence can contain at least $2^{64}-2^{63} = 2^{63}$ boxes. We define no mechanism for extending sequences to more than $2^{63}$ boxes, but applications could use a box to communicate a new read capability if such an extension were required. 

Both Alice and Bob can now derive a sequence of KDF symmetric keys $H_i$, location blinding factors $K_i$, symmetric payload encryption keys $E_i$, box IDs $M_i$, and increment $i\rightarrow i+1$. The $i$ is used as additional input in the $H_i$ KDF as control input to reduce the risk of incorrect implementations generating valid KDF outputs for incorrect $i$ indices.

$$H_i,\ i  \underset{\text{KDF}}{\rightarrow} \ H_{i+1}, \ E_i, \ K_i.$$
Both parties generate the message encryption key $E_i^{\text{ ctx}}$:
$$E_i \ ,\ \text{ctx}  \underset{\text{KDF}}{\rightarrow}  E^{\text{ ctx}}_i,$$
and the blinding factor $K_i^{\text{ ctx}}$, and the blinded public key $M_i^{\text{ ctx}}$ which will serve as the \emph{box ID}:
\begin{align*}
K_i \ ,\ \text{ctx} &  \underset{\text{KDF}}{\rightarrow}  K_i^{\text{ ctx}}, \\
M_i^{\text{ ctx}} & =P_R \cdot K_i^{\text{ ctx}}, \\
& \equiv B \cdot S_R \cdot K_i^{\text{ ctx}}.
\end{align*} 

\noindent Alice derives a $M_i^{\text{ ctx}}$-specific secret scalar $S_i^{\text{ ctx}}$, and encrypts message $m_i$ as ciphertext $c_i^{\text{ ctx}}$ using key $E_i^{\text{ ctx}}$, and signs it as $s_i^{\text{ ctx}}$ using $S_i^{\text{ ctx}}$. This signature ensures unforgeability (\textrm{EUF-CMA}) from adversaries that possess a read capability for the sequence, enabling the use of BACAP in group settings with multiple readers:
\begin{align*}
c_i^{\text{ ctx}} & = \textrm{AES-256-GCM-SIV-ENCRYPT}(m_i \ , \ E_i^{\text{ ctx}}), \\
S_i^{\text{ ctx}} & = S_R \times K_i^{\text{ ctx}} \pmod{\ell}, \\
s_i^{\text{ ctx}} & = \textrm{Ed25519-SIGN}(c_i^{\text{ ctx}} \ , \ S_i^{\text{ ctx}}).
\end{align*}

\noindent Alice sends her \emph{message} to the server:
\vskip -0.1 in

$$M_i^{\text{ ctx}} , \ c_i^{\text{ ctx}}, \ s_i^{\text{ ctx}}.$$

\noindent The server verifies the \emph{write capability} to ensure it was sent by a sequence \emph{writer}, as opposed to a \emph{reader}.
$$ \textrm{Ed25519-VERIFY}(M_i^{\text{ctx}}, \ c_i^{\text{ctx}}, \ s_i^{\text{ctx}})$$

\noindent Bob requests $M_i^{\text{ ctx}}$ from the server, verifies and decrypts:$$ \textrm{Ed25519-VERIFY}(M_i^{\text{ctx}}, \ c_i^{\text{ctx}}, \ s_i^{\text{ctx}}),$$
$$m_i=\textrm{AES-256-GCM-SIV-DECRYPT}(c_i^{\text{ctx}},E_i^{\text{ ctx}}). $$
The original Ed25519 \cite{ed25519} signing algorithm works on \emph{private keys} that are not scalars, but SHA-512 hash preimages created as part of the signing operation. We refer to a modified Ed25519 signing algorithm that skips the hash step and instead operates directly on private scalars as \textrm{Ed25519-SIGN}.
\medskip

\subsection{Choice of $\textbf{encrypt}$/$\textbf{decrypt}$ functions} 
Using an authenticated symmetric encryption scheme prevents a \emph{third-party} quantum adversary from forging $c_i^{\text{ctx}}$, separate from the $s_i^{\text{ctx}}$ signature, the private key for which is obtained by the adversary. Such an adversary can forge signatures over garbled ciphertexts, but not plaintexts, and can't authenticate the ciphertexts. We use an authenticated cipher scheme, \textrm{AEAD\_AES\_256\_GCM\_SIV} \cite{rfc8452} (using $M_i^{\text{ ctx}}$ as \emph{nonce}), to make implementation less error-prone, and to enable replacing a $c_i$ with a \emph{tombstone}, a term we will define in \autoref{pigeonhole-tombstones}.

\medskip

\subsection{Forward-security} BACAP 
achieves \emph{computational post-quantum forward-security} by the irreversibility of the KDF function. Parties may choose to 
leverage this by throwing away state associated with $H_{i-1}$ once done with it. Instead of sending the \emph{read capability} at $i_0$, Alice may choose to instead reveal a later $\{ P_R, H_{i_n}, i_n\}$ which would only enable a newcomer Charlie to read the sequence starting from $i_n$. We sample the original $i_0$ randomly instead of starting at $0$, to avoid indirectly revealing to Charlie how many boxes preceded $i_n$.

We separate $E_i$ and $K_i$, as Alice may want to keep $K_i$ to be able to derive $S_i^{ \text{ctx}}$ at a later point, for example to sign a message to delete the box, without being able to decrypt the message payload encrypted under $E_i^{ \text{ctx}}$. Separating the two permits a protocol instantiation to selectively have forward-security only for $E$ and the resulting $c$ ciphertexts.

The knowledge of $S_i^{ \text{ctx}} \equiv S_R \times K_i^{ \text{ctx}} \pmod{\ell} $ and the factor $K_i^{ \text{ctx}}$ makes it trivial to recover $$S_R \equiv S_i^{ \text{ctx}} \times (K_i^{ \text{ctx}})^{-1} \pmod{\ell}.$$
As a result, sharing the derived signing keys with third-parties is not safe because the recipient can recover the effective signing key for the whole sequence.

\subsection{Unlinkability}

 \noindent Suppose that an adversary $\mathcal{A}$ knows the values in BACAP available to the server. Let $X$ be the event that box IDs $M$ and $M'$ belong to the same sequence $\{M_i^{ctx}\}_{i}$, and $X'$ that they don't. Let $X_\mathcal{A}$ be the event that the adversary guesses that they do. Then $M$ and $M'$ are \emph{unlinkable} iff $$|\mathbb{P}[X_\mathcal{A}|X]-\mathbb{P}[X_\mathcal{A}|X']|\leq \delta, $$ for some sufficiently small $\delta\geq 0.$

It is not possible for anyone without \emph{read capabilities} to determine whether two messages in different boxes belong to the same sequence. The symmetric encryption of $m_i$ under key $E_i^{ \text{ctx}}$ (unknown to adversaries) results in \emph{unlinkability} even against \emph{chosen-plaintext attacks}.

The messages in BACAP remain unlinkable to a quantum adversary.
While the security of the Ed25519 signing scheme, and thus the distinction between BACAP \emph{read} and \emph{write} capabilities, relies on the hardness of the elliptic curve discrete logarithm problem (ECDLP),\footnote{Solved with Shor's algorithm on a quantum computer.} the \emph{unlinkability} does not:

Solving ECDLP for $B$ in $M_i = B \cdot S_R \cdot K_{i}^{ \text{ctx}}$ yields a scalar $S_{R} \times K_{i}^{ \text{ctx}} \in \mathbb{F}_{\ell}$ (the effective signing key) for each box $M_i^{ \text{ctx}}$.
An adversary who knows $S_R$ or a $K_i^{ \text{ctx}}$ can trivially compute the modular multiplicative inverse, but because $S_R$ and $K_i^{ \text{ctx}}$ are independent pseudo-random elements  of $\mathbb{F}_{\ell}$, there is no unique solution to the equation $M_i^{ \text{ctx}} = S_R \times K_i^{ \text{ctx}} \pmod{\ell}$ that lets the adversary solve for $S_R$. Such a solution would enable them to link $M_{x}^{ \text{ctx}}, M_{y}^{ \text{ctx}}$.  
Therefore, the \emph{post-quantum unlinkability} of BACAP relies on:
\begin{enumerate}[itemsep=0ex, parsep=0ex, parsep=0ex]
\item The indistinguishability of $K_i^\text{ ctx} \pmod{\ell}$  from uniformly random elements  of $\mathbb{F}_{\ell}$  with negligible bias. That is, an assumption that the KDF is secure, and that the reduction $\pmod{\ell}$ has negligible bias.
\item The computational difficulty of enumerating the $K_i^{ \text{ctx}}$ (keyspace roughly $2^{256}$).
\item The property that for any $S_R$ guess, enumeration of $H_0, \text{ctx} \rightarrow K_x^{ \text{ctx}}, K_y^{ \text{ctx}}$ is required to find two $K$s such that $M_x^{ \text{ctx}} = B \cdot S_R \cdot K_x^{ \text{ctx}}$ and $M_y^{ \text{ctx}} = B \cdot S_R \cdot K_y^{ \text{ctx}}$.
\end{enumerate}

\medskip

\subsection{Post-compromise security} A basic implementation of BACAP does not provide post-compromise security. A simple way to achieve post-compromise security would be to rotate BACAP sequences frequently. It is worth noting that deriving new $H_n, i_n$ from a KDF (without communicating a new $S_R$/$P_R$) provides \textit{post-compromise security} with respect to \emph{unlinkability} since the adversary would still not be able to link $K_i$s or obtain $E_i$s. 

A quantum adversary can impersonate writes by solving ECDLP for $S_R \times K_i^{ \text{ctx}} \pmod{\ell}$ (but still not obtain new $K_i^{ \text{ctx}}$), so nothing new is learned. In the classical setting, adversary knowledge of a compromised $S_R$ does not help the adversary obtain the $K_i^{ \text{ctx}}$ required to compute $S_R \times K_i^{ \text{ctx}} \pmod{\ell}$.


\section{Pigeonhole storage}\label{pigeonhole-section}


The goal of this section is to extend our system to provide messaging functionality expected by today's Internet users without compromising our security goals, and to futher strengthen our resistance to service nodes being compromised. 
This is achieved with Pigeonhole servers, which provide a time-limited storage capacity to implement asynchronous, unidirectional, single-writer, multi-reader BACAP messaging channels with strong metadata protection for clients against both passive and active adversaries, including authorized readers who are malicious or compromised. 
As Alice exchanges messages with Bob, they rely on pseudorandom shared sequences of BACAP boxes as storage locations. Boxes in the same sequence are linkable to users with read or write capabilities for the sequence, but are cryptographically unlinkable to the storage servers thanks to the properties of BACAP. 


\subsection{Additional concerns in interactive messaging}

\subsubsection{Increased vulnerability to statistical disclosure} Messaging comes with a number of statistical pitfalls. 
These include user behavior being vulnerable to correlation, the number of queries to particular boxes revealing that they belong to the same BACAP sequence with the same number of readers, and the difference in time between writes and reads. In the case of two users being more likely to be online at the same time if they are talking to each other, little can be done unless users themselves are mindful of the correlation potential and able to adjust their habits. In the case of other correlations, \emph{pigeonhole services} employ a number of powerful mitigating tactics.

\subsubsection{Reliability vs forced interactivity and lossiness}
A mix network can be expected to 
drop a small number of in-flight messages to load-shed \cite{seda}. For applications requiring reliable delivery, this necessitates a reliability mechanism. There is a tension between reliability and metadata protection due to the information that is revealed by sending acknowledgements or retransmitting missing information. Expecting a user to acknowledge receipt of messages might make her vulnerable to confirmation attacks and correlation. 
Some mixnet-based systems \cite{nym} automatically retransmit unacknowledged messages. This occurs transparently (to the user/sender) and repeats until acknowledgement is received. Others, like Karaoke\cite{karaoke} acknowledge the risk of repeated events, and terminate the users' conversation if message loss is detected.



 The tension applies not only to the short-term reliability needs of ensuring that a write from a user to a service node was received, but also to the long-term reliability problem where a writer needs to retransmit older writes based on the their perception of the reader's state. In \autoref{couriers} we address both facets of this problem with \textit{couriers}, which are aware of retransmits but are not aware of the box IDs they are associated with, and we're able to provide reliability without automatic acknowledgements. 

\subsubsection{Long-term channel resilience} If a BACAP message becomes unavailable before the intended recipient(s) retrieves it, the recipient may continue trying to read the message forever. Messages can become unavailable either due to server failure, or as part of scheduled garbage collection to make room for new messages to be stored. To ensure \emph{eventual consistency}, we will introduce a mechanism to redeliver messages with \emph{all-or-nothing backfills}.

\subsection{All-or-Nothing Design Philosophy}

Suppose a user sends two or more related messages. If each message can fail independently, and the adversary (such as the contact) observes some messages appearing, it alerts them to the intention of sending additional messages. The failure of these messages to be observed can be correlated with a physical connectivity issue.
We follow a principle that actions should either \emph{succeed completely} or \emph{fail unobservably}, and describe our solutions to these problems in \autoref{reliable-group-channels}.

\subsection{Replicas and sharding}


To further mitigate the risks of service nodes in a messaging application being compromised, we split their functionality between \emph{storage servers} or \emph{replicas}, and \emph{couriers}. Couriers maintain fixed-throughput connections to replicas, as do replicas with each other.\smallskip

\begin{figure}[ht!]
\centering
\begin{tikzpicture}[scale=1.15,shorten >=1pt,node distance=3cm,on grid,auto]
\node [cloud, black, draw,cloud puffs=10,cloud puff arc=120, aspect=0.5, inner ysep=2em, fill=purple!20!white, opacity=0.35] at (-2,1) {};

\draw[black] (0,-0.5) node {\footnotesize COURIERS};
\draw[black] (2.5,-0.5) node {\footnotesize REPLICAS};
\draw[black] (-2,-0.2) node {\footnotesize GATEWAYS AND };
\draw[black] (-2,-0.5) node {\footnotesize MIX NODES};

\node[client, minimum size=0.5cm, black] (client0) at (-4,0.5) {\footnotesize BOB};
\node[client, minimum size=0.5cm, black] (client1) at (-4,2) {\footnotesize ALICE};
\node[draw=none] (clientjoin) at (-1.5,1) {};

\foreach \L in {0,1,2}{
  \node[server, fill=purple!50!blue, minimum size=0.85cm] (courier\L) at (0,\L) {
  };  }

\node[server, fill=blue!50!black, minimum size=0.85cm] (replica0) at (2.5,0) {
  };
\node[server, fill=blue!50!black, minimum size=0.85cm] (replica1) at (1.5,1) {
  };
\node[server, fill=blue!50!black, minimum size=0.85cm] (replica2) at (2.5,2) {
  };

\foreach \L in {0,1,2}{
\foreach \M in {0,1,2}{
\draw (replica\L) edge[<->,purple!20!white,opacity=0.35,midway,right, ultra thick] (courier\M);
}}

\node[server, fill=blue!50!black, minimum size=0.85cm] at (1.5,1) {
  };

\draw[<->, ultra thick, green!50!black] (-3.75,2) -- (-2.5,1.5);
\draw[<->, ultra thick, red!70!black] (-3.75,0.45) -- (-2.5,0.5);
\draw[<->, ultra thick, green!50!black] (-1.45,0.7) -- (-0.35,0);
\draw[<->, ultra thick, red!70!black] (-1.45,1.2) -- (-0.35,2);

\foreach \L in {0}{
  \draw (courier\L) edge[<->,green!50!black, ultra thick] (replica0);
  \draw (courier\L) edge[<->,green!50!black, ultra thick] (replica1);
};

\draw (courier0) edge[<->,green!50!black, ultra thick] (replica1);
\draw (courier2) edge[<->,red!70!black, ultra thick] (replica2);

\draw (replica1) edge[<->,blue!60!black,midway,right, ultra thick] (replica0);
\draw (replica2) edge[<->,blue!60!black,midway,right, ultra thick] (replica0);
\draw (replica1) edge[<->,blue!60!black,midway,right, ultra thick] (replica2);
\label{fig:pigeonhole-component-diagram}
\end{tikzpicture}
\caption{Replication is marked in \textcolor{blue!50!black}{blue}, Alice's write operation \textcolor{green!50!black}{green}, Bob's read operation \textcolor{red!70!black}{red}, and Couriers' fixed-throughput connection to the replicas in \textcolor{purple!60!black}{purple}.}
\end{figure}
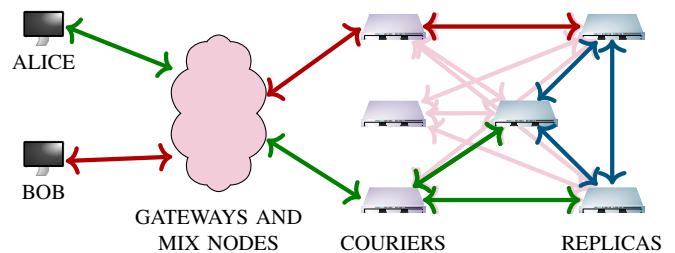

Suppose we have $n$ pigeonhole storage servers, or \emph{replicas}, and we want a subset of $k$ of them to store each box. The storage system is sharded using a consistent hashing \cite{consistent_hashing} scheme which allows entities with knowledge of a the box ID (BACAP's $M_i^{\text{ctx}}$) to deterministically select the two servers that are currently responsible for that box. The $k$ servers responsible for storing a given box are then that box's \emph{replicas}. 



The consistent hashing method makes this efficient. For a given box ID, one derives a set of $k$ servers by sorting the list of servers by a hash of their public key concatenated with the box ID and use the result to select a permutation of the set of $n$ servers for each box ID. We can then choose the first $k$ of them. This results in a set of $k$ designated replicas indistinguishable from being chosen uniformly at random.
When a storage server goes offline, or a new server joins only proportion of $k/n$ boxes need to be retransmitted.







\subsection{Couriers} \label{couriers}

The second type of service, couriers, can be seen as a mix network service layer and are responsible for acknowledging the client's requests, but do not learn the box ID. They communicate with the replicas over fixed-throughput direct connections outside the mix network and avoid revealing the existence of a retransmission to the storage server.

Commands sent through couriers are encrypted by the client to the target replicas' NIKE keys, which are published in the PKI and rotated each epoch to provide forward secrecy. This encryption can optionally be post-quantum if the NIKE key is a hybrid NIKE as is used in our PQ NIKE Sphinx construction described in \autoref{pq-nike-sphinx}.


\subsubsection{Writing messages}
To store a BACAP message in the network, Alice first generates an ephemeral NIKE keypair which will be used for the envelope, and a symmetric encryption key ("envelope key") which will be used to hide the envelope contents from the courier. She then randomly picks a \emph{courier}, and two \emph{intermediate replicas} that will receive write operation from the courier. The \emph{intermediate replica} that receives an envelope decrypts it, commits it to disk, and acknowledges to the courier that the replica takes responsibility for delivering the message to the final replicas. The courier then sends an acknowledgement to the client.

These \emph{intermediate replicas} are chosen independently of the two \emph{final replicas} for that box ID which are derived using the sharding scheme. 
The reason Alice designates \emph{intermediate} replicas, as opposed to addressing the \emph{final} replicas directly, is to avoid revealing to the courier which shard the box falls into. In a system with many replicas, this partitioning would otherwise allow a courier to collude with one of Alice's contacts to execute SURB floods.


\subsubsection{Reading messages}

To read from a box for the first time, a client generates an ephemeral keypair for a new \emph{read envelope}. It selects one of the replicas responsible for that box according to the sharding scheme, and creates a \emph{read request} containing the ephemeral public key, the replica ID, the box ID encrypted with the ephemeral private key and the replica's public key. It chooses a random courier, and sends it the read request.

Upon receiving a read request, the courier forwards the read envelope to the designated replica.
who decrypts the envelope, and checks if it has an entry in its database for the referenced box ID. 
If an entry does exist, the replica encrypts a reply with the BACAP message encrypted to the envelope public key. It also encrypts a reply encrypted to the envelope public key, but containing a negative acknowledgement instead of the desired BACAP message. The nature of these responses should be indistinguishable to the courier.

If the box was empty but is written to in the near future, the replica schedules new replies to each of the listeners. Each reply is independently delayed, with delays sampled from a uniform distribution to mitigate the courier's ability to infer links between responses that pertain to the same box. 

The delay also serves to hide relationships between writers and readers, to avoid a pending read from being fulfilled immediately upon the courier sending a write envelope to a replica. This mitigation is needed to address the cases where writer/reader or reader/reader pairs are using the same courier, but ultimately it does not address all concerns, which we will elaborate on in \autoref{pigeonhole-security}. The courier caches a listener response associated with a particular envelope for a time.


A client that receives a negative acknowledgement may poll the box again in the near future. To prevent informing replicas of the precise rate of read requests sent by a client, which could link client behavior information with the box ID, the client sends the same read envelope to the same courier, and the courier will refrain from re-sending that envelope to the replica more than once. When the courier sees a duplicate envelope, it uses the new SURB associated with the recent read request to repeat the replica's latest response for that envelope. 

Couriers for retried read requests are rotated frequently to mitigate correlation attacks where the courier could otherwise link a number of concurrent readers ceasing to poll at the same time and infer that they were interested in the same box.

\subsubsection{Tombstones} \label{pigeonhole-tombstones}

BACAP messages come with a fixed-size payload $c_i$.
To allow users to delete messages after sending them, we selectively break the unlinkability guarantees provided by BACAP with \emph{tombstones}, which are BACAP messages with empty $c_i$. 
When a replica gets a tombstone for an $M_i$ that it has an existing $c_i,s_i$ for, the tombstone takes precedence and the replica deletes the old $c_i,s_i$ pair.
Tombstones 
are also used in \autoref{reliable-group-channels} for end-to-end reliability.


\subsection{Server context and storage duration}

BACAP provides for adding \textit{blinding contexts} used when deriving keys, and in the Echomix PKI. \cite{katzenpki}
provides a Shared Random Value \cite{sharedrandom}
and a set of previous \emph{Weekly Shared Random Values} (WSRV). Clients use the last WSRV as a blinding context so that the box addresses they are querying rotate, and the period for which a Pigeonhole server is able to observe query patterns for a particular box is bounded. 

Pigeonhole storage servers collect boxes in per-week buckets, and discard the oldest bucket upon entering a new week. A minimum guaranteed storage time can be calculated from the total storage capacity and the maximum total possible network throughput.

\subsection{End-to-end reliable group channels} \label{reliable-group-channels}
Sometimes a reader is offline longer than the replicas' data retention period, or the two replicas responsible for a box both fail. We therefore need an end-to-end reliability mechanism. This requires some form of acknowledgement from readers to writers and retransmissions from writers to readers, with a user knowingly disclosing information to contacts in these rare cases. To make sure a malicious writer can't probe whether a reader of their channel is online, acknowledgements are never sent automatically. Instead they are sent opportunistically with the next message the acknowledging user sends on their own channel. Likewise, the retransmission operation is never performed automatically, but only when the user sends a message.



We perform retransmissions in an all-or-nothing fashion by introducing a new courier \emph{copy} command, 
 to prevent a malicious reader from learning about network disruptions which affect a targeted writer while they are 
 retransmitting a series of messages.
To write or rewrite multiple messages to a channel, the writer creates a new pigeonhole channel to temporarily store the encrypted write requests for writing to the boxes which it ultimately wants to write to. 
After completing all of the writes to the temporary channel, it sends a \emph{copy} command to a random courier, which contains the write capability for the temporary channel. The courier derives the read capability for it, reads the encrypted write commands from it, and executes each as it would normal write commands. The new message which the user wanted to send, which triggered the retransmission operation, will be the last write operation in the temporary channel.

The courier does not learn the box IDs of the boxes in the long-term channel it is writing to, because the write operations are encrypted to the replicas chosen by the client. The courier uses the temporary channel's write cap to write tombstones to delete the copied temporary boxes.

These end-to-end reliable channels can be thought of as a single extremely resilient multi-writer channel, which enables group communication applications which are robust against the loss of all data stored by replicas. Communication can even be resumed on a whole new set of Pigeonhole servers without users needing to re-bootstrap their channels.

\subsection{Future work and variations on the design}

\subsubsection*{PIR} Both read and write requests provide replicas with a rough, probabilistic ordering of the box IDs, and the guesses get better with each reader. At the cost of efficiency, this could be addressed with a private information retrieval (PIR) system to protect read operations. 
PIR could streamline our reliable channel mechanism: each channel could have a single \emph{recovery box} in a PIR system where they can write one message per SRV to catch readers up to a point where they can find our messages, instead of writing and reading a large number of tombstones or old messages to reestablish communication with a contact who has been offline for a long time.

\subsubsection*{Push SURBs} Absent a PIR scheme, there are variations of Pigeonhole where read requests include \emph{two} SURBs: one for immediate acknowledgement of the request, and another to be used later to send the payload after the write subsequently happens. To preserve receiver unobservability with respect to the gateway, this may also necessitate \emph{long echoes}: decoy messages containing two SURBs, one of which is used in the future to provide inbound cover traffic after the user has disconnected.

\subsubsection*{SURB burning} Using the echo service to intentionally invalidate (unreceived) SURBs previously sent to couriers, to limit exposure to SURB floods.

\section{Security properties of Pigeonhole messaging over Echomix}\label{pigeonhole-security}

The analysis in this section assumes an Echomix network where each service node is exclusively running a Pigeonhole courier, and an echo service for echo decoys, without a use of unrelated services. 

We consider several active network roles, and the effects of collusion between the combinations of two of them, as well as a GPA. Our goal is that collusion between any two of these active roles is insufficient to meaningfully compromise metadata confidentiality. However, an adversary who can compromise some combinations of two or more of these roles is able to perform some useful attacks. 

\subsection{Single-role adversary capabilities}


\subsubsection*{Global Passive Adversary} Learns which clients are connecting to the mixnet, and learns the traffic patterns between all elements of the network.
\textbf{In all following adversary descriptions, we assume GPA capabilities implicitly.}

 \subsubsection*{Contact (reader)} When they see a message, the reader can infer that the writer was online within the period of time defined by the latency parameters of the network. This can enable an intersection attack, singling out the clients that were connected around the time of each write.

 \subsubsection*{Gateway} Sees the incoming and outgoing packets, but can't distinguish between decoys and pigeonhole messages.

\subsubsection*{Replica}  Learns which box IDs are being written to the replica, and when.
        Learns which box IDs are being read, and when.
        Can withhold envelope replies temporarily or indefinitely.
        Can lie about having messages (either sending bogus responses or denying having stuff they have received).
        Learns which courier was responsible for a read.

\subsubsection*{Courier} 
     Learns the rate of resends for a given enveloped write message, and that they come from the same client.
     Can withhold responses temporarily or indefinitely. Withholding a response results in the client eventually retrying the send. The courier can link these, and could use them in correlation attacks, but the courier alone does not know which boxes the envelopes are related to.
     Can drop envelopes or reads, denying a client the services they requested, but cannot target a speciic client.



\subsection{Capabilities of colluding pairs of compromised network elements}

\noindent We assume GPA capabilities. \emph{Mixes} assumes all three intermediate mix nodes on the path are compromised.

\subsubsection*{Gateway $+$ Mixes} Since this adversary cannot distinguish echo traffic from courier traffic, and service nodes are picked at random, this does not yield useful information. 
  %


\subsubsection*{Gateway $+$ Courier} A courier receiving a copy request can accumulate SURBs as the client retries, and send them in a burst. At the gateway, it can see this burst and link a user to that copy request. This adversary knows the length of the backfill operation, but not the destination box IDs.


\subsubsection*{Gateway $+$ Replica} Although the courier pinning mitigates it somewhat, this adversary does get to perform a long-term intersection attack by observing which clients are connected when certain boxes are being read.

\subsubsection*{Gateway $+$ Contact} A gateway and contact can do a long-term intersection attack, as the contact knows in which epoch a user sent a message. This adversary can more efficiently confirm if the user is a client of the gateway by dropping or delaying users messages and observing when messages are received by the contact, when the contact is using the compromised gateway.

\subsubsection*{Mixes $+$ Courier} Similar to the capabilities of \emph{Gateway, Courier}, except they link the operations to a specific gateway rather than a specific user


\subsubsection*{Mixes $+$ Replica} No useful attacks.

\subsubsection*{Mixes $+$ Contact} No useful attacks.

\subsubsection*{Courier $+$ Replica} Learns which boxes are being written and read. Coupled with observations of a user population and their network disruptions (GPA) can identify the client writing to, and clients reading from, a specific box, when both reader and writer pick the compromised courier+replica combination. Can link box IDs in backfills involving the compromised replica.

\subsubsection*{Courier $+$ Contact} No useful attacks.

\subsubsection*{Replica $+$ Contact} Knows when boxes are read/written. For 1:1 conversations, they learn when the other party was online, even though they were just doing a read. For groups, they learn when \emph{someone} in the group was reading, but can't necessarily distinguish readers from each other.

\section{Post-quantum security}\label{section-pq}

We have added quantum-resistant cryptographic primitives in a hybrid scheme that combines their guarantees with those of elliptic curve cryptography. These are all implemented in golang cryptographic libraries which are generalised and accessible to other projects. \cite{hpqc}

\begin{enumerate}[itemsep=0ex, parsep=0ex, parsep=0ex]
\item The consensus producing the PKI document, uses the post quantum hybrid signature scheme of Ed25519 \cite{ed25519} combined with Sphincs+\footnote{Not to be confused with the Sphinx packet format.} \cite{sphincs}. 

\item We use an upgraded Noise protocol \cite{noise, noiseanalysis}, PQNoise, as described in \cite{pqnoise}. It uses KEM \footnote{KEM: key encapsulation mechanism} \cite{kem} as opposed to the EC Diffie-Hellman in Noise. Katzenpost \cite{kp} is the first implementation of PQNoise. 

\item We have two hybrid post-quantum updates to the Sphinx packet format \cite{sphinx}. One that adds cryptographic agility to the classical Sphinx so that it can use any NIKE \footnote{NIKE: non-interactive key exchange}. We then use the hybrid post quantum NIKE of X25519\cite{curve25519} combined with CTIDH \cite{ctidh}. The other is a KEM--based nested encryption packet format. 
It can likewise use any KEM. It is significantly faster, but has larger packet headers.



\end{enumerate}

\subsection{Post-quantum mixnet packets}

We will now elaborate on the updates to Sphinx. The Echomix/Katzenpost packet encryption has two interchangeable ways to achieve post quantum security. 

\subsubsection{Post-quantum NIKE Sphinx} \label{pq-nike-sphinx}

Our implementation of NIKE Sphinx uses a generic set of NIKE \cite{nike} interfaces that allow any NIKE, adding cryptographic agility to classic Sphinx. We use a hybrid NIKE consisting of CTIDH512 and X25519. This comes at a relatively high computational cost, and so it is appropriate for latency-tolerant implementations with a lower messaging frequency. It preserves the compactness of classic Sphinx by using its \emph{blinding trick} \cite{sphinx}. 

As in classic Sphinx, the body plaintext contains an integrity tag and is nested encrypted with an SPRP\footnote{SPRP: strong pseudo-random permutation, or a wide-block cipher.} as the payload $\delta$, while the header is composed of three parts:

\begin{itemize}[itemsep=0ex, parsep=0ex, parsep=0ex]
    \item $\alpha:$ A NIKE public key,
    \item $\beta:$ Symmetrically encrypted routing information section,
    \item $\gamma:$ A MAC.
\end{itemize}

Suppose we have a mix node $n$, with private key $x_n$. It transforms the Sphinx packet by replacing $\alpha,\ \beta,\ \gamma$ with $\alpha',\ \beta',\ \gamma'$. The Sphinx blinding trick lets the client compose several NIKE public keys where each key is generated by a node from the last one using the blinding operation. In particular, we generate $\alpha'$: $$\alpha, x_n \underset{DH}{\longrightarrow} S, $$ $$\alpha,b(S)\underset{blind}{\longrightarrow} \alpha'. $$

A shared secret $S$ is computed using the packet header's public key and the mix node's private key. A KDF is used to generate several other secrets, including a blinding factor $b(S)$. $\alpha'$ is computed by blinding $\alpha$ with $b(S)$. And so we don't need to include separate public keys for different hops, but we are doing additional calculations.

Other operations performed by the node are as follows: it will use $S$ to compute a hash of $\beta$, and compare it to $\gamma$ to verify the integrity of the header. Then it will strip a layer of encryption from the payload, and obtain $\beta',\ \gamma':$ $$\beta,\ p(S)\underset{\oplus}{\longrightarrow} \beta',\ \gamma',\ n',$$ where $n'$ is the identity of the next node. It will then send off $\alpha',\ \beta',\ \gamma'$ and the payload to $n'$.

This is a straightforward implementation of Sphinx with added cryptographic agility. We will now compare it to KEM Sphinx.


\subsubsection{Post-quantum KEM Sphinx}

We will now introduce KEM Sphinx, our KEM-based \cite{kem}, Sphinx revision which uses a generic set of KEM interfaces. Similar to our NIKE Sphinx, KEM Sphinx is meant to be used with a hybrid post quantum KEM in order to achieve post quantum security. A good default choice could be Xwing \cite{xwing}. However our implementation makes available a more general purpose way to compose PQ KEMs using a generic secure KEM combiner that lets one combine an arbitrary number of KEMs while preserving IND-CCA2 security if at least one of the underlying KEMs has IND-CCA2 security. \cite{kemcombiners}

The KEM ciphertexts are stored in the $\beta$ section of the header, which makes it significantly larger. They are nested encrypted, and the original stream cipher \textit{xor}ed padding scheme is obeyed. In the routing slot for each hop, the first element is always the KEM ciphertext. The packet still consists of the following parts:
\begin{itemize}[itemsep=0ex, parsep=0ex, parsep=0ex]
    \item $\alpha:$ A KEM ciphertext,
    \item $\beta:$ Symmetrically encrypted routing information and KEM public keys,
    \item $\gamma:$ A MAC,
    \item $\delta:$ The packet's payload.
\end{itemize}

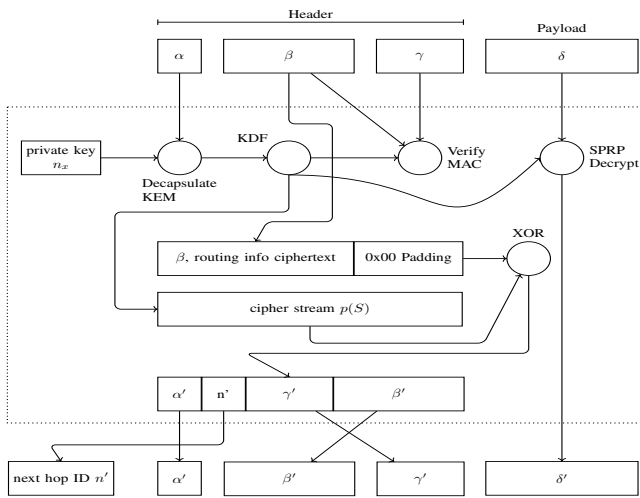
\begin{figure}[!ht]
\centering
\resizebox{8.5cm}{6.5cm}{%
\begin{tikzpicture}[scale=1]
  \tikzset{
    symbol/.style={rectangle, draw, minimum height=1cm, align=center},
    circleNode/.style={circle, draw, minimum size=1cm, align=center}
  }

  \node[symbol, minimum width=1cm] (alpha) {$\alpha$};
  \node[symbol, minimum width=3cm, right=0.5cm of alpha] (beta) {$\beta$};
  \node[symbol, minimum width=2cm, right=0.5cm of beta] (gamma) {$\gamma$};
  \node[symbol, minimum width=3.5cm, right=0.5cm of gamma, label=above:Payload] (delta) {$\delta$};

  \node[circleNode, below=2cm of alpha, label={[align=left]below:Decapsulate\\KEM}] (decap) {};
  \node[circleNode, below=2cm of beta, label=north west:KDF] (kdf) {};
  \node[circleNode, below=2cm of gamma, label={[align=left]right:Verify\\MAC}] (mac) {};
  \node[circleNode, below=2cm of delta, label={[align=left]right:SPRP\\Decrypt}] (sprp) {};

  \node[symbol, minimum width=1cm, left=1.3cm of decap] (newRect) {private key\\$n_x$};

  \draw [|-|] ([yshift=.5cm]alpha.north west) -- ([yshift=.5cm]gamma.north east) node[midway, above] {Header};

  \draw[->] (newRect) -- (decap);
  \draw[->] (alpha) -- (decap);
  \draw[->] (gamma) -- (mac);
  \draw[->] (delta) -- (sprp);
  \draw[->] (decap) -- (kdf);
  \draw[->] (kdf) -- (mac);

  \draw[->] (kdf.south) to [out=0, in=230, looseness=1.7] (sprp.west);
  
  \node[symbol, minimum width=4.5cm] (routing info) [below=6cm of alpha.north west, anchor=north west] {$\beta$, routing info ciphertext};
  \node[symbol, minimum width=2.5cm, right=0cm of routing info] (padding) {0x00 Padding};

  \node[circleNode, label=north:XOR, right=1cm of padding] (xor) {};

  \node[symbol, minimum width=7cm] (cipher stream) [below=7.5cm of alpha.north west, anchor=north west] {cipher stream $p(S)$};
  
  \node[symbol, minimum width=1cm, below=9cm of alpha] (alpha2) {$\alpha'$};
  \node[symbol, minimum width=1cm, right=0 of alpha2] (next hop) {n'};
  \node[symbol, minimum width=2cm, right=0 of next hop] (gamma2) {$\gamma'$};
  \node[symbol, minimum width=3cm, right=0 of gamma2] (beta2) {$\beta'$};

  \draw[->,rounded corners] (beta.south) |- ++ (1,-1.5) |- ++ (-1.5,-2.9) -- (routing info.north);

  \draw[->,rounded corners] (kdf.south) |- ++(-4,-1.1) |- (cipher stream.west);

  \draw[->] (padding.east) -- (xor.west);
  \draw[->,rounded corners] (cipher stream.south) |- ++(4, -.5) -- (xor);
  \draw[->,rounded corners] (xor.south) |- ++(-6.5, -2.3) -- (gamma2.north);

  \node[symbol, minimum width=1cm, below=11.5cm of alpha] (alpha3) {$\alpha'$};
  \node[symbol, left=1cm of alpha3] (next hop2) {next hop ID $n'$};
  \node[symbol, minimum width=3cm, below=1.5cm of gamma2] (beta3) {$\beta'$};
  \node[symbol, minimum width=2cm, right=.5cm of beta3] (gamma3) {$\gamma'$};
  \node[symbol, minimum width=3.5cm, below=11.5cm of delta] (delta2) {$\delta'$};

  \draw[->] (sprp.south) -- (delta2.north);
  \draw[->] (alpha2) -- (alpha3);

  \draw[thick,dotted] ($(decap.north west)+(-3.6,1.15)$) rectangle ($(sprp.south east)+(1.6,-7.5)$);

  \draw[->] (beta2) -- (beta3);
  \draw[->] (gamma2) -- (gamma3);
  \draw[->,rounded corners] (next hop.south) |- ++(-4, -1) -- (next hop2.north);

  \draw[->] (beta) -- (mac);

\end{tikzpicture}
}
\caption{A circuit diagram of unwrapping a KEM Sphinx message $((\alpha, \beta, \gamma), \delta)$
into $((\alpha', \beta', \gamma'), \delta')$ at mix $n$.}
\end{figure}


 Since we do not use the blinding trick, the header is less compact than that of classic Sphinx. It is most appropriate when a given usage is able to compensate for the header overhead by making the packet payload bigger. 

 
$$\alpha, x_n \underset{\text{decap}}{\longrightarrow} S, $$ 
$$\beta,\ p(S)\underset{\oplus}{\longrightarrow} \alpha',\ \beta',\ \gamma',\ n',$$

\subsubsection{Speed}

Unwrapping KEM Sphinx packets is roughly twice as fast than the classical NIKE Sphinx since it involves one public key operation rather than two. We no longer calculate the group element for the next hop by blinding the current group element. Instead, we extract the new KEM ciphertext from the encrypted routing information section of the Sphinx packet header. The following table compares the header size to Sphinx unwrap speed on a 11th Gen Intel(R) Core(TM) i7-1165G7 @ 2.80GHz processor for Sphinx variants.

{\renewcommand{\arraystretch}{1.5}
\begin{table}[h!]
\centering
\begin{tabular}{ |c|c|c|c| } 
 \hline
\footnotesize\textbf{Sphinx type} &\footnotesize\textbf{resistance} & \footnotesize\textbf{ns/op} & \footnotesize\textbf{header size}\\ 
 \hline
\scriptsize X25519 NIKE  & \scriptsize ECC & \scriptsize 151,383 & \scriptsize 476 \\
\scriptsize X448 NIKE & \scriptsize ECC & \scriptsize 254,966 & \scriptsize 500 \\
\scriptsize X25519 KEM & \scriptsize ECC & \scriptsize 57,611 & \scriptsize 636 \\
\scriptsize X448 KEM & \scriptsize ECC & \scriptsize 208,326 & \scriptsize 780 \\
\scriptsize Xwing KEM & \scriptsize hybrid& \scriptsize 175,732 & \scriptsize 7,164 \\
\scriptsize MLKEM768-X25519 KEM &\scriptsize hybrid& \scriptsize 182,334 & \scriptsize 7,164 \\
 \hline
\end{tabular}
\caption{Sphinx variants speed in nanoseconds per operation, and header size in Bytes. It should be pointed out that X25519 KEM Sphinx is nearly three times as fast as the standard X25519 NIKE Sphinx, but the header is only one third larger.}
\end{table}
}
\normalsize

\subsubsection{Related work} Another approach to using KEM with nested encryption packets can be found in the recently published EROR packet format \cite{eror}. However, this format assumes doubling the payload overhead. 
The goal of this is to add protection from a tagging attack, in which an adversary who controls both a node on the way and a service can corrupt the payload ciphertext and link the packet with a packet that arrives at a service when it doesn't decrypt. However, since the adversary learns nothing else about the packet this way, and the information gained on other packets from this is negligible, we propose that in the context of a mixnet like Echomix this is not a practical trade-off.
Since a practical implementation of KEM nested encryption may benefit from offsetting the large header size with a large payload size, doubling the payload size appears to be additionally costly. 

\section{Latency and bandwidth overhead}

If a packet's journey is comprised of $k$ steps, each incurring a mean delay $\mu=\frac{1}{\lambda}$, then their sum will follow the Erlang distribution, $$f_{k,\lambda}(x)=\frac{\lambda^k x^{k-1}e^{-\lambda x}}{(k-1)!}.$$ Since a single delay has both mean and standard deviation equal to $\mu$, the sum will quickly approach a normal distribution with mean $k\mu$ and standard deviation $\sigma=\sqrt{k}\mu$: $$\underset{k\rightarrow \infty}{\lim}\frac{\lambda^k x^{k-1}}{(k-1)!}e^{-\lambda x} = \frac{\lambda}{\sqrt{k} 2\pi}e^{-\frac{\lambda^2}{2k}((x-1/\lambda)^2}.$$

In the Echomix design the round trip involves 9 steps (gateway - 3 mix nodes - service - 3 mix nodes - gateway), not including the client's initial sending scheduler, which is a separate parameter. The round-trip latency obeys the Erlang distribution for $k=9$: $$f_{9,\lambda}(x)=\frac{\lambda^9 x^{8}e^{-\lambda x}}{8!},$$ with cumulative distribution function: $$F_{9,\lambda}(x)=1- \sum_{n=0}^8\frac{1}{n!}e^{-\lambda x}(\lambda x)^n. $$ This means that the total expected round-trip latency is $9\mu$, and the probability of exceeding $20\mu$ is about $0.002.$

For example, in the Echomix deployment by Zero Knowledge Network \cite{0kn} 
the average delay per hop is $\mu=0.2$s, resulting in average round-trip latency of 1.8s, and a $0.2\%$ chance of exceeding 4s. 

The client bandwidth use is a function of the packet size and send frequency, and cryptographic primitives used, and the size of the PKI document which the clients download from gateways. 
Apart from the PKI document, which is downloaded at most every 20 minutes, the client's overhead is independent of the number of nodes in the mix network.
The size of the PKI document is linear in the number of nodes in the network. 

In the example of $\emptyset$ Knowledge Network, each packet's user payload size is 30kBs, using an X25519 NIKE Sphinx with an additional 1kB of header and SURB size. With an average of 2.5 packets sent per second, the clients send and receive about 77kBs, which means that a client connected continuously will send and receive about 6.7GBs of data per day, with up to 96\% of this memory being usable payloads. These parameters are practical for messaging, medium bitrate audio transmission and interfacing with many internet services such as cryptocurrency blockchains.

{\renewcommand{\arraystretch}{1.5}
\begin{table}[h!]
\centering
\begin{tabular}{ |c|c|c| } 
 \hline
\footnotesize\textbf{Sphinx type} &\footnotesize\textbf{resistance} & \footnotesize\textbf{header $+$ SURB size}\\ 
 \hline
\scriptsize X25519 NIKE  & \footnotesize ECC &  \footnotesize 1,082 \\
\scriptsize X448 NIKE & \footnotesize ECC &  \footnotesize  1,130\\
\scriptsize CTIDH1024 NIKE & \footnotesize post-quantum &  \footnotesize  2,030\\
\scriptsize CTIDH1024-X448 NIKE & \footnotesize hybrid &  \footnotesize  3,226\\
\scriptsize X25519 KEM & \footnotesize ECC &  \footnotesize  1,402\\
\scriptsize X448 KEM & \footnotesize ECC &  \footnotesize  1,690\\
\scriptsize Xwing KEM & \footnotesize hybrid&  \footnotesize  14,458\\
\scriptsize MLKEM768-X25519 KEM &\footnotesize hybrid&  \footnotesize  14,458\\
\scriptsize MLKEM768-X448 KEM &\footnotesize hybrid&  \footnotesize  14,746\\
 \hline
\end{tabular}
\caption{Per-packet bandwidth overhead in Bytes on a mix network with a round-trip of 9 hops for different Sphinx variants. A comprehensive list can be found in Appendix IV.}
\end{table}
}
\normalsize

The PKI document grows with the size of the network, since it has to include each node's public key information. This document is typically downloaded once per 20 minute epoch. The directory authorities can also produce a smaller document detailing changes from the previous epoch, so that clients only need to download the full PKI document when they first connect.

{\renewcommand{\arraystretch}{1.5}
\begin{table}[h!]
\centering
\begin{tabular}{ |c|c|c|c|c| } 
 \hline
\footnotesize\textbf{Sphinx} & \footnotesize\textbf{dirauths} & \footnotesize\textbf{nodes} & \footnotesize\textbf{replicas} &  \footnotesize\textbf{size}\\ 
 \hline
 \scriptsize X25519  & \footnotesize 3 & \footnotesize  10 & 0 & \footnotesize  159,901 \\
 \scriptsize X25519  & \footnotesize 9 & \footnotesize  10 & 0 & \footnotesize  459,421 \\
 \scriptsize X25519 & \footnotesize 3 & \footnotesize  500 & 5 & \footnotesize  167,076 \\
  \scriptsize X25519 & \footnotesize 9 & \footnotesize  500 & 100 & \footnotesize  1,071,855 \\
  \scriptsize CTIDH1024-X448  & \scriptsize 3 & \footnotesize  10 & 0 & \footnotesize  159,901 \\
 \scriptsize CTIDH1024-X448  & \scriptsize 9 & \footnotesize  10 & 0 & \footnotesize  459,421 \\
 \scriptsize CTIDH1024-X448 & \scriptsize 3 & \footnotesize  500 & 5 & \footnotesize  167,836 \\
 \scriptsize CTIDH1024-X448 & \scriptsize 9 & \footnotesize  500 & 100 & \footnotesize  1,087,055 \\
 \hline
\end{tabular}
\caption{Example size of the PKI document in Bytes. Directory authorities are assumed to use Ed25519 and Sphincs+ hybrid signatures, and replicas are assumed to use Xwing.}
\end{table}
}
\normalsize

\section{Conclusion}

The Echomix mix network design surpasses the state of the art systems by providing stronger metadata privacy and resistance to global adversaries who compromise users and parts of the network infrastructure, and eliminating multiple vulnerabilities of previously published systems. In particular, it uses symmetry and unobservable traffic coupling to meaningffully protect against traffic analysis, avoiding the mistakes of its predecessors. Echomix is implemented as a robust real-world open source software project, Katzenpost \cite{kp}, elements of which have been used by several systems, including Zero Knowledge Network, Cloaked Services and our own chat client, Katzen.

In order to extend rigorous security guarantees to the difficult case of persistent multi-user messaging, we introduce the blinding-and-capability (BACAP) cryptographic protocol, which allows users to unlinkably interface with pseudorandom nodes. 
Pigeonhole storage, in conjunction with BACAP, provides additional powerful privacy properties and extends the messaging protocol to include functionality expected by today's users, such as both short and long term reliability and deleting messages, while maintaining the anonymity-first design philosophy. Together, and implemented on top of Echomix, they are suitable for low-latency, interactive group messaging in the presence of realistic adversaries. It is the first practical messaging system design with such strong threat model.


We also introduce KEM Sphinx and cryptographic agility to the Sphinx packet format, achieving hybrid post-quantum security in both KEM and NIKE Sphinx. KEM Sphinx is faster than NIKE Sphinx, but carries more bandwidth overhead.

Finally, we demonstrate that this design is practical, efficient, with manageable latency and bandwidth overhead and can yield a comfortable user experience for many of today's Internet services.

\section*{Acknowledgment}


This work was funded by a research grant from the Wau Holland Foundation and by Zero Knowledge Network. Echomix/Katzenpost development is funded by the Wau Holland Foundation, EU Horizon2020 grant ID: 653497, NLNet, Protocol Labs, Zero Knowledge Network.

\bibliographystyle{unsrt}
\bibliography{sources}

\newpage

\section*{Appendix I: Machine-checkable proofs}

For the reader's convenience, we provide the machine-checkable proofs in their original electronic format, ready for verification. They can be found at \href{https://github.com/katzenpost/research/}{https://github.com/katzenpost/research/}. along with the full lists of Sphinx geometry overheads and PKI document sizes.
\smallskip



\section*{Appendix II: Structure of courier read/write requests} \label{courier-commands}
\noindent A Sphinx\cite{sphinx} payload destined for a courier contains (for \emph{read} and \emph{write} requests):\medskip

\noindent \textbf{Shared fields:}
\indent \begin{enumerate}[itemsep=0ex, parsep=0ex, topsep=0ex]
\item The sender's ephemeral hybrid public key:
  \begin{enumerate}[itemsep=0ex, parsep=0ex, parsep=0ex]
    \item \textrm{x25519} public key
    \item \textrm{CTIDH-1024} public key
  \end{enumerate}
\item for each designated replica: \begin{enumerate}[itemsep=0ex, parsep=0ex, topsep=0ex]
  \item 256-bit DEK encrypted to the replica's public key
 \end{enumerate}
\item enveloped message which the courier can't decrypt
\end{enumerate}
\medskip For \textbf{write requests, couriers see:}
\indent \begin{enumerate}[itemsep=0ex, parsep=0ex, topsep=0ex]
    \item \textbf{(shared fields)}
    \item SURB for courier to ACK receipt of request once at least one replica has accepted it
\end{enumerate}
For \textbf{read requests, couriers see:}
\begin{enumerate}[itemsep=0ex, parsep=0ex, topsep=0ex]
    \item \textbf{(shared fields)}
    \item Immediate-use SURB, for courier to ACK receipt of the encrypted (to the client) reply from the designated replica
    \item replica/shard id designating each replica to contact
\end{enumerate}

\section*{Appendix III: Structure of replica envelopes} \label{appendix-replica-envelopes}

\medskip \noindent Replicas see \textbf{write requests} as:
\begin{enumerate}[itemsep=0ex, parsep=0ex, topsep=0ex]
\item sender's ephemeral public key
\item envelope DEK encrypted with shared secret between sender private key and replica public key
\item enveloped message, encrypted with DEK, containing a BACAP message:
  \begin{enumerate}[itemsep=0ex, parsep=0ex, topsep=0ex]
    \item BACAP box ID ($M_i^{\text{ ctx}}$)
    \item BACAP payload ($c_i^{\text{ ctx}}$)
    \item BACAP signature ($s_i^{\text{ ctx}}$)
  \end{enumerate}
\end{enumerate}

\noindent Replicas see \textbf{read requests} as:
\begin{enumerate}[itemsep=0ex, parsep=0ex, parsep=0ex]
\item sender's ephemeral public key
\item envelope DEK encrypted with shared secret between sender private key and replica public key
\item enveloped message, encrypted with DEK, containing a BACAP box ID:
  \begin{enumerate}[itemsep=0ex, parsep=0ex, parsep=0ex]
    \item BACAP box ID ($M_i^{\text{ ctx}}$)
  \end{enumerate}
\end{enumerate}



\end{document}